# Ultrafast many-body dynamics of dense Rydberg gases and ultracold plasma

Check for updates

**Mario Großmann[1], Jette Heyer[1,2,3,5], Julian Fiedler[1,2,3,5], Markus Drescher[1,3,4], Klaus Sengstock[1,2,3], Philipp Wessels-Staarmann[1,2,3] ✉ & Juliette Simonet[1,2,3]**

Understanding Coulomb driven many-body dynamics in ultracold atomic systems far from equilibrium remains an open challenge, particularly when ultrafast excitation channels create competing pathways toward Rydberg gases or ultracold plasmas. Here, we investigate the many-body dynamics in a $^{87}$Rb Bose-Einstein condensate after exposure to a single femtosecond laser pulse. By tuning the laser wavelength across the two-photon ionization threshold, we can control the initial state that is either dominated by free electrons and leads to an ultracold plasma or dominated by electrons in excited states which leads to a dense Rydberg gas. The large bandwidth enables overcoming the Rydberg blockade that limits the excitation density for narrow-linewidth lasers. We directly measure the kinetic energy of the released electrons and compare the final distribution of free, bound and plasma electrons to molecular dynamics simulations where the electrons are modeled as individual particles including collisional ionization and recombination processes. We find very good agreement between the simulated electron distribution and the experimental observation. We identify charge imbalance as main driver for the decay of a dense Rydberg gas into an ultracold plasma.

Ultrashort laser pulses can induce, control, and probe exotic phases of matter in atomic[1,2], molecular[3–5], cluster[6], and solid-state systems[7–11]. Understanding the many-body dynamics and the effect of electronic correlations is of increasing importance for the interpretation of these experiments[12–15].

The broad spectral bandwidth of ultrashort laser pulses allows exciting or ionizing thousands of ultracold atoms by a single laser pulse[16,17] to investigate many-body phenomena dominated by Coulomb interactions, such as the formation of ultracold plasma or Rydberg gases on ultrafast time-scales. For example, picosecond laser pulses are used to observe the dephasing of electronic coherence in an ultracold Rydberg gas by time-domain Ramsey interferometry[18] or to excite ultracold atoms into Rydberg states in a Mott insulator, where a suppression of Penning ionization is evident if the Rydberg orbitals do not overlap[19]. Femtosecond laser pulses in combination with ultracold quantum gases allow exploiting strong-field ionization at high atomic densities of a Bose-Einstein condensate (BEC), which gives rise to an initially strongly coupled plasma[20] that leads to ultrafast cooling of the electrons.

In strongly coupled plasma, the Coulomb energies dominate the thermal energies, which leads to fascinating phenomena such as spatial self-organization[21,22]. These systems are potential models for dense astrophysical objects like white dwarfs and neutron stars[23], warm dense matter[24], and for understanding transport processes for nuclear fusion[25].

So far, experimental approaches to realize strongly coupled plasma have focused on photoionized ultracold atoms in magneto-optical traps[26,27] or molecules in supersonic expansions[28] which leads to ultracold neutral plasma with negligible charge imbalance. In such experiments, the coupling strength is limited by disorder-induced heating[27,29,30], which can be minimized by ion cooling[31] or magnetic trapping of the plasma[32]. Alternative approaches to introduce an initial spatial correlation by using optical lattices or a blockaded Rydberg gas have also been investigated[33–35].

The question on the stability of Rydberg gases has triggered many theoretical and experimental studies to understand the dynamics resulting from ionizing, recombination, and state-changing collisions between the constituents[36–42] as well as spreading of Rydberg excitations[43–45]. It has been reported that a dense Rydberg gas can decay into a plasma through Penning ionization and subsequent electron-Rydberg collisions, while three-body recombination converts an ultracold plasma into a Rydberg gas[28,46–48].

Open questions regarding this transition remain: How stable are dense Rydberg gases with overlapping electron orbits? Is it possible to create a plasma with negligible electron energy out of Rydberg excitations to reach the strong-coupling regime? Is there a difference between an ionized plasma with vanishing excess energy and a plasma forming from a Rydberg gas with vanishing binding energy? Is there an equilibrium between three-body-recombination and ionizing collisions?

[1]Center for Optical Quantum Technologies, University of Hamburg, Hamburg, Germany. [2]Institute for Quantum Physics, University of Hamburg, Hamburg, Germany. [3]The Hamburg Centre for Ultrafast Imaging, University of Hamburg, Hamburg, Germany. [4]Institute of Experimental Physics, University of Hamburg, Hamburg, Germany. [5]These authors contributed equally: Jette Heyer, Julian Fiedler. ✉e-mail: philipp.wessels-staarmann@uni-hamburg.de

Here, we report on the investigation of the many-body dynamics triggered by a femtosecond laser pulse in ultracold $^{87}$Rb gases with adjustable densities up to a BEC. By tuning the femtosecond laser wavelength across the two-photon ionization threshold, we can control the excess energy of the electrons from positive values in the continuum, which leads to the formation of an ultracold plasma dominated by a charge imbalance, to negative values associated with the excitation of bound states, which results in the realization of a dense ensemble of Rydberg atoms. The latter regime is experimentally inaccessible by continuous wave lasers as the Rydberg blockade[49,50] limits the density of Rydberg excitations for narrow-bandwidth lasers[51,52].

Our experimental setup allows to probe the electron kinetic energies and distinguish between free, bound and plasma electrons. We compare the measured final composition of the many-body system to complete molecular dynamics simulations, which treat the electrons as individual particles and go beyond the usual approach of modeling the electrons as a background density. Experimentally realized systems below a few thousand excited electrons allow including all microscopic ionization and recombination processes in the simulation and enable a precise comparison to the experimental results.

We find a very good agreement between measurement and simulation over a broad range of parameters and are able to reconstruct the ultrafast dynamics taking place within the first pico- to nanoseconds, which explains the final electron composition. The dynamics reported here differ significantly from the behavior of ultracold neutral plasma, as an initial charge imbalance hinders the three-body recombination of low-energy electrons into Rydberg states and favors the decay of electrons in excited states into an ultracold plasma.

## Results

### Ultrafast excitation and ionization of a quantum gas

By exposing an $^{87}$Rb BEC to a femtosecond laser pulse with tunable wavelength in the visible range, an ultracold gas consisting of ions, electrons, and atoms in excited states can be created on ultrashort time-scales by multiphoton excitation/ionization. Whether the system is dominated by ionization or excitation of bound states can be tuned by adjusting the excess energy

$$E_{\mathrm{exc}} = 2 \times E_{\mathrm{ph}} - E_{\mathrm{b}} \tag{1}$$

of the electrons, which accounts for the additional energy released by simultaneous absorption of two photons of tunable photon energy $E_{\mathrm{ph}}$. Moreover, the excess energy depends on the (positive) binding energy $E_{\mathrm{b}}$ of the electron in the ground state atom. Thus, ionization happens at positive excess energies, while negative energies lead to excitation of Rydberg and bound states:

$$E_{\mathrm{exc}} > 0 \Rightarrow \text{ free electron}$$
$$E_{\mathrm{exc}} < 0 \Rightarrow \text{ bound electron / Rydberg electron}$$

At high excess energies, a fraction of the free electrons of the ionized system rapidly escapes the ionization volume and leaves behind a positively charged region where the remaining initially free electrons become trapped in a highly charged microplasma with large electron trajectories around the ion core (see Fig. 1a). The trapped electrons are referred to as plasma electrons for which ultrafast electron cooling on pico- to nanosecond time scales has been observed[20]. Here, the kinetic energy of the electrons is transferred to the potential energy of the ions and reducing the excess energy leads to a more neutral plasma (Fig. 1b) with smaller electron trajectories.

At the ionization threshold, as shown in Fig. 1c, both photoionization and excitation of Rydberg states can occur due to the large bandwidth of the femtosecond laser pulse. Here, the interplay between the Penning or collisional ionization of atoms in Rydberg states and three-body-recombination becomes relevant.

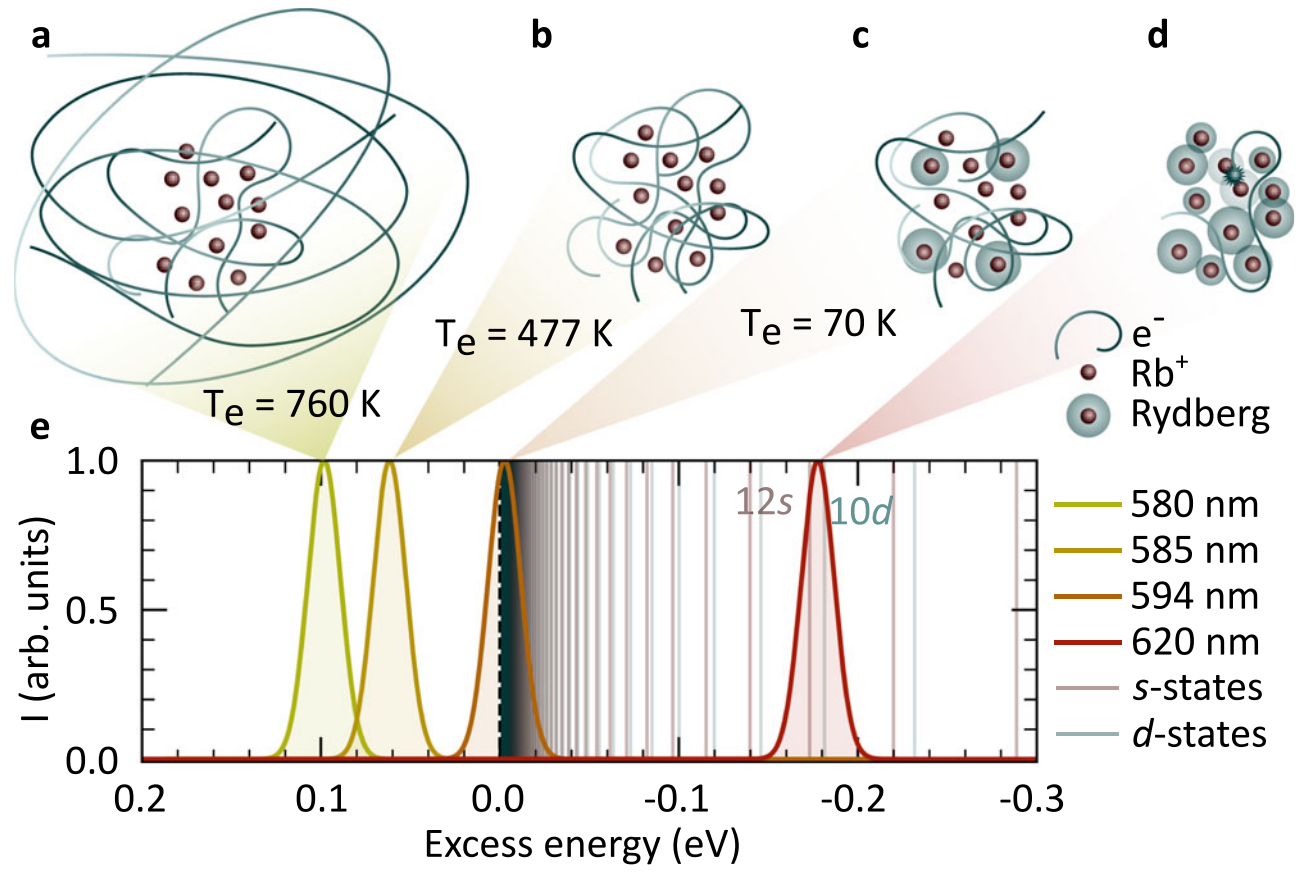

**Fig. 1 | From ultracold plasma to dense Rydberg gases.** Manipulating a Bose-Einstein condensate (BEC) with a femtosecond laser pulse can lead to a variety of different systems: By reducing the excess energy of electrons, the dynamics can be tuned from that of a highly charged microplasma (**a**) with large electron orbits around the ionic cloud to more neutral plasmas (**b**) close to the ionization threshold. For vanishing excess energy (**c**), Rydberg atoms are created due to the large bandwidth of the laser pulse and Rydberg recombination can occur. For negative excess energies (**d**), the system enters the regime of Rydberg gases with Penning ionization due to overlap of Rydberg orbitals. **e** The excitation spectra for two-photon processes at wavelengths ranging from 580 to 620 nm are plotted for a pulse duration of 170 fs. The vertical lines mark the ionization threshold (black dotted line) as well as the *s*-states of $^{87}$Rb (red) and the *d*-states (green) which can be addressed.

Tuning the excess energy to negative values, the two-photon process can no longer lead to photoionization but can resonantly excite various *s*- and *d*-states of $^{87}$Rb from the 5*s* ground state. The investigated system will be a Rydberg gas whose stability will strongly depend on density and binding energy (see Fig. 1d). In Fig. 1e, the excitation spectra for a two-photon process are illustrated for the experimentally accessible parameters.

### Experimental realization

Figure 2a illustrates the experimental setup. An untrapped $^{87}$Rb BEC (see "Methods") is exposed to a single laser pulse with a duration of 166(4) fs (see "Methods"). The pulse is focused down to a waist of $w_0 = 1$ μm by means of a high numerical aperture microscope objective allowing peak intensities on the order of $10^{13}$ W/cm$^2$. The focused femtosecond laser beam is smaller than the extent of the atomic cloud so that the ultracold atoms are ionized locally (see "Methods"). Ions and electrons emerging from ionization processes are separated by antisymmetric potentials at opposing extraction meshes and spatially resolved by combining a microchannel plate (MCP) and phosphor screen recorded by a high-speed camera (see Supplementary Note 1). The population of bound states can be accessed by detecting the photoelectrons after applying a 1064 nm Rydberg ionization pulse.

During the propagation of the electrons towards the detector, kinetic energy is translated into spatial information. As depicted in Fig. 2b, we use charged particle tracing (CPT) simulations to reconstruct the spatial distribution corresponding to kinetic energies in a range between 0.01 and 2 eV. The non-centrosymmetric distributions obtained in our experimental setup are due to the grounded shielding mesh below the microscope objective (see Fig. S1 in Supplementary Note 1). A comparison between simulations and experimental images allows us to distinguish different energy classes of electrons.

Using an optical parametric amplifier (OPA), we tune the wavelength of the femtosecond laser pulse across the two-photon ionization (2PI) threshold at $\lambda_{2\mathrm{PI}} = 593.7$ nm in a range between 580 and 640 nm (see Fig. 2c).

At positive excess energies according to Eq. (1), where the energy of two photons from the femtosecond laser exceeds the binding energy, photoelectrons with kinetic energies $E_{\mathrm{kin,2PI}} < 0.1$ eV are emitted. On resonance with the ionization continuum at vanishing excess energy, the bandwidth of the two-photon process $\sigma_{2\gamma}$ = 9.1 meV FWHM becomes relevant with contributions both above and below the threshold due to the short pulse duration. At negative excess energies, the two-photon processes can no longer lead to ionization, but can excite *s* and *d* Rydberg states. In this regime, the large bandwidth of the laser mitigates blockade effects (see Fig. S3 in Supplementary Note 4) and allows the creation of dense Rydberg gases.

At high peak intensities, the simultaneous absorption of three photons from the laser pulse leads to three-photon ionization (3PI) at negative excess energies or above threshold ionization (ATI) at positive excess energies, yielding more free electrons with high kinetic energies of 1.6 eV < $E_{\mathrm{kin,3PI}}$ < 2.2 eV in the given wavelength range.

In addition to the directly emitted electrons, we can detect atoms in Rydberg states by using a 10 µs laser pulse at $\lambda_{\mathrm{Ry,PI}}$ = 1064 nm for photoionization (see "Methods"). By subtracting the average images with and without this additional photoionization pulse, the signal corresponding to the electrons from ionized Rydberg states can be extracted. Rydberg states with binding energies up to 0.4 eV then correspond to detected electrons with kinetic energies 0.8 eV < $E_{\mathrm{kin,Ry}}$ < 1.2 eV.

Figure 2d illustrates this process: The left grayscale image shows the measured kinetic energy distribution on the electron detector after applying a femtosecond laser pulse on a BEC which addresses bound states below the ionization continuum at negative excess energies. The image is composed of three electron classes:

The small black dot is associated with low-energy plasma electrons as a result of the energy transfer from the electrons to the ions in the plasma dynamics[20]. The remaining signal is a superposition of electrons with high kinetic energy originating either from 3PI or ATI processes (referred to as free or 3PI electrons) or from bound and Rydberg states that have been ionized by the additional photoionization pulse at $\lambda_{\mathrm{Ry,PI}}$ (referred to as bound or Rydberg electrons). Subtracting the average image without the additional ionization pulse (center grayscale image) then isolates the Rydberg electrons (right image).

## From ultracold plasma to dense Rydberg gases

As our experimental setup allows measuring the electron kinetic energy, we can separate between free, bound and plasma electrons. Thus, we can probe the electron composition at the end of the many-body dynamics with respect to the excess energy that sets the initial composition of the system. This allows realization of ultracold microplasma with free charge carriers at positive excess energies. At negative excess energies a dense Rydberg gas with a significant admixture of atoms in bound states will form.

Figure 3a shows the measured electron composition for positive and negative excess energies across the 2PI threshold (gray dotted vertical line). The plot indicates the recorded detector brightness (shaded areas show the standard deviation) as well as the derived electron number $N_e$ (see Supplementary Note 2). The Rydberg signal (red) is extracted as explained in the previous section (compare Fig. 2d) and cold plasma electrons (green) and electrons originating from 3PI (yellow) are separated according to their kinetic energy. The total electron signal as the sum of the three components is shown as dark gray curve. The mask used for separating plasma and 3PI electrons is exemplified on a mean detector image in Fig. 3b. In addition to the measurements for a BEC (top), we also perform measurements with a dilute thermal cloud (bottom), which correspond to a density that is approximately 570 times lower (see "Methods").

Three central observations can be made regarding the electron signal:

- The Rydberg signal (red) shows distinct resonances for a dilute cloud, while the signal appears less pronounced for a BEC.
- There is a significantly increased occurrence of 3PI (yellow) when ionizing an atomic target at BEC densities compared to the dilute cloud, especially below the 2PI threshold.
- We find electrons associated with plasma formation (green) at all excess energies for a BEC and dilute cloud, even at wavelengths where the two-photon process is far below the ionization threshold and resonant to deeply bound states.

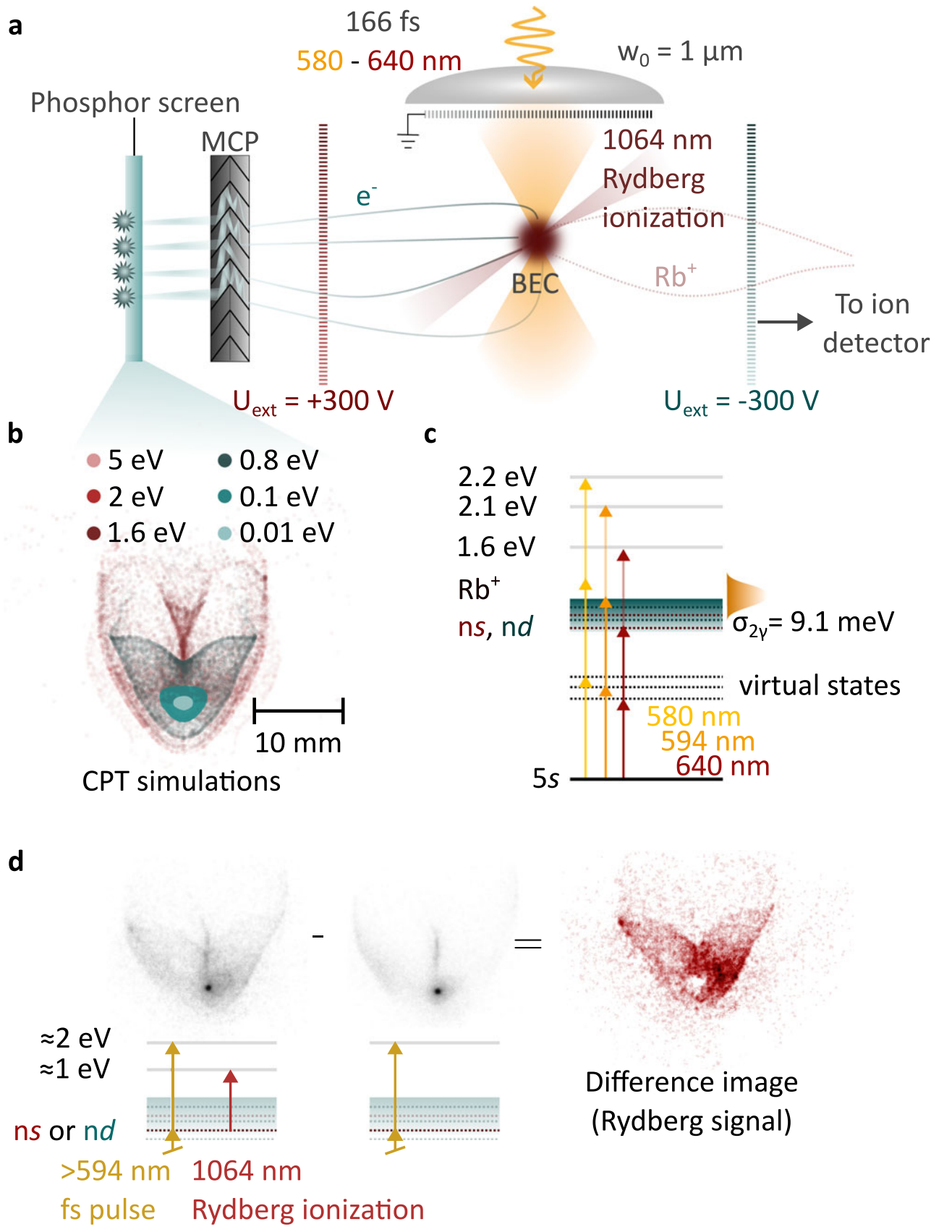

**Fig. 2 | Energy-resolved detection of electrons emerging from ultrafast ionization of a Bose-Einstein condensate (BEC). a** A fraction of an $^{87}$Rb BEC is ionized by a focused femtosecond laser pulse with tuneable wavelength ranging from 580 to 640 nm. The emerging electrons and ions are separated by an extraction field and spatially resolved by two microchannel plates (MCPs) and phosphor screens. The population of bound states is accessible by detecting the photoelectrons after a 1064 nm Rydberg ionization laser pulse. The kinetic energy of the electrons is translated into spatial information during their propagation towards the detector. Comparing the recorded detector image with charged particle tracing (CPT) simulations (**b**) of monoenergetic electron distributions allows to estimate the maximum kinetic energy in a range between 0.01 and 2 eV. **c** Two- and three-photon ionization of $^{87}$Rb at wavelengths of 580, 594, and 640 nm. At the threshold, the bandwidth of the two-photon process is $\sigma_{2\gamma}$ = 9.1 meV FWHM. Below the threshold, the two-photon process yields ultrafast excitation of *s*- and *d*-states. **d** To detect Rydberg states excited at wavelengths below the ionization threshold, a Rydberg ionization pulse at $\lambda_{\mathrm{Ry,PI}}$ = 1064 nm is used, yielding photoelectrons with a kinetic energy of approximately 0.8−1.2 eV, depending on the binding energy of the Rydberg state. The difference of the mean images with and without this ionization pulse shows a clear signature of these electrons.

The Rydberg electrons detected in the dilute cloud (red) display distinct resonances close to the unperturbed 12*s*/10*d*, the 11*s*/9*d* and the 10*s*/8*d* state, and their width is consistent with the bandwidth of two-photon processes $\sigma_{2\gamma}$ = 9.1 meV (FWHM). The latter two peaks make up most of the total electron signal.

The number of detected electrons populating bound states in the vicinity of the 10*s*/8*d* state after excitation within a single femtosecond laser pulse in a BEC reaches approximately 50% of the detected electrons. For such deeply bound states this corresponds to an excitation density of $10^{19}$ m$^{-3}$, which is stable during the plasma dynamics. The large bandwidth of the femtosecond laser pulse makes it possible to excite two neighboring atoms to Rydberg states up to $n \approx 30$, even at the density corresponding to a BEC, as blockade effects are far less relevant than in experiments with continuous wave Rydberg excitation schemes (see Fig. S3 in Supplementary Note 4). This would allow realizing an ultracold Rydberg gas with an excitation density beyond the experimentally reported $\approx 5 \times 10^{16}\,\mathrm{m}^{-3}$ for $n = 38$[18].

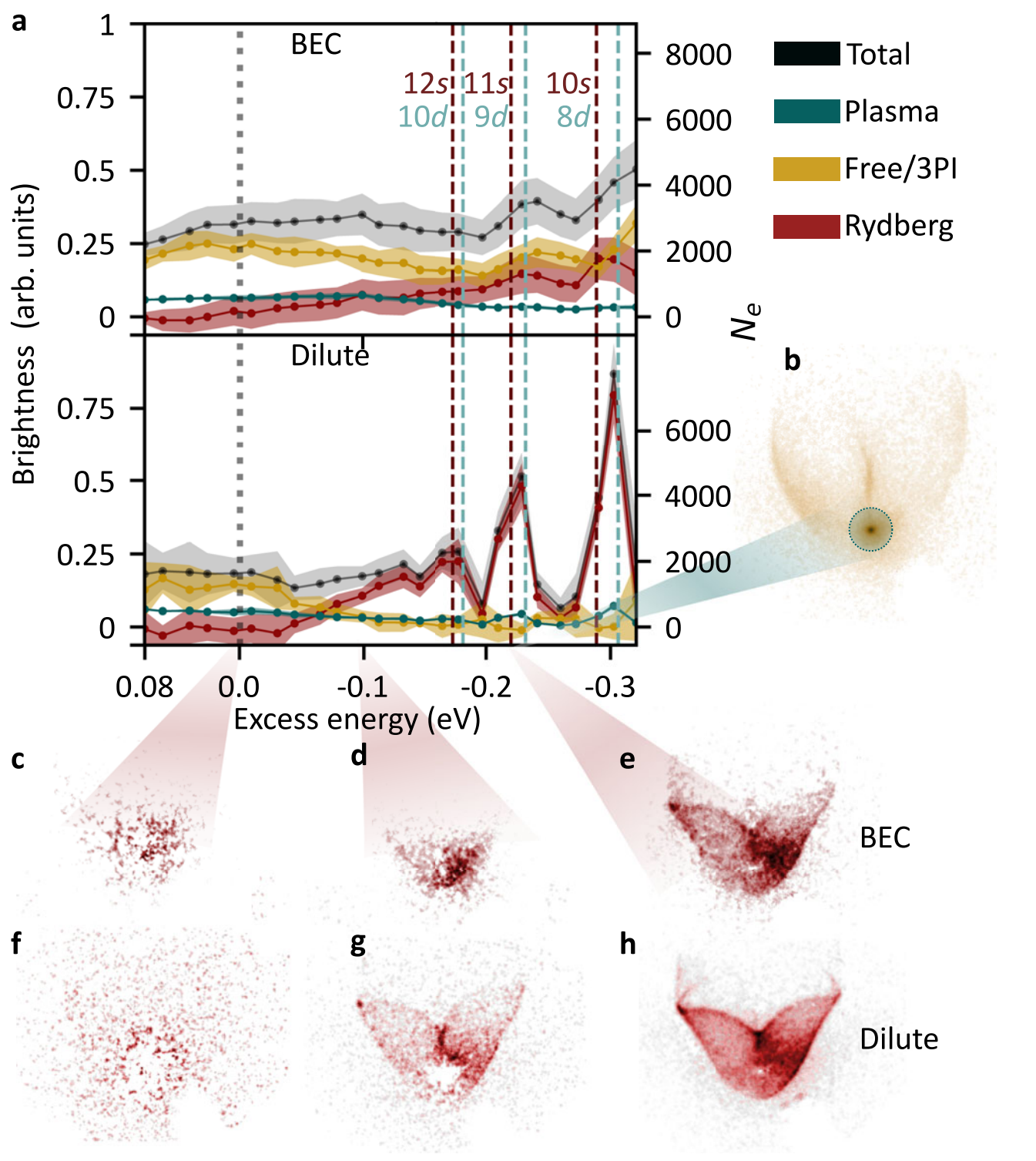

**Fig. 3 | Electron composition of dense Rydberg gases and ultracold plasma.** **a** Measured detector brightness and corresponding electron numbers $N_e$ separated into plasma (green), Rydberg (red) and free/three-photon ionization (3PI) (yellow) electrons as a function of the excess energy for a Bose-Einstein condensate (BEC) (top) and a dilute thermal cloud (bottom). The sum of the three electron components is depicted as dark gray curve. The ionization continuum is indicated as dotted gray line. Shaded areas show the standard deviation on the detector brightness. **b** Exemplary mean detector image without Rydberg ionization pulse. The signal marked yellow in the image is accounted to free/3PI electrons (compare to Fig. 2b at 2 eV). The dotted green circle shows the mask used to extract the plasma electrons with kinetic energy smaller than 0.1 eV. Normalized detector images of the kinetic energy distribution of the Rydberg electrons for a BEC exposed to femtosecond laser pulses at $E_{exc} = 0$ eV (**c**), at $E_{exc} = -0.1$ eV (**d**), and at $E_{exc} = -0.23$ eV (**e**). **f**–**h** Normalized detector images of the kinetic energy distribution of the Rydberg electrons for a dilute cloud at the same excess energy values than for a BEC.

Despite the significantly lower density, more bound and Rydberg excitations are detected in the dilute cloud due to a larger ionization volume compared to the BEC. This results from the larger overlap between the intensity distribution of the femtosecond laser beam and the thermal cloud compared to the smaller BEC (see "Methods"). Moreover, the lower density of the thermal cloud leads to a more stable Rydberg population due to less likely Penning ionization (see Fig. S2 in Supplementary Note 3) compared to the measurement with a BEC, where the peaks appear less pronounced.

Note that for experimental runs with very small to no Rydberg atoms the number of Rydberg electrons (red) in Fig. 3a may take negative values as this is a difference signal between two detector images as described in the previous section (compare Fig. 2d). The negative values lie within the standard deviation of the Rydberg signal (red shaded area) so that negative values are consistent with a vanishing number of Rydberg electrons.

Figure 3c–h exemplarily shows the measured kinetic energy distribution of the electrons after photoionization of the Rydberg states at three different excess energies for a BEC in the upper row and for a dilute thermal cloud in the bottom row. Note that each detector image here is normalized for illustrating the shape on the detector. For two-photon processes resonant with the 11*s* and 9*d* states, the obtained energy spectra for a thermal cloud (Fig. 3h) and for a BEC (Fig. 3e) are both consistent with the expected kinetic energy of the electrons after photoionization of the Rydberg state. Closer to the ionization continuum, the kinetic energy distributions for a dilute cloud and a BEC are significantly different. Here, the data suggest that the reduced density of the dilute target in comparison to the BEC increases the stability of larger Rydberg states during the dynamics, since Penning ionization depends on the amount of overlap of the electron orbitals[42] and ionization triggered by collisions with electrons in the plasma fraction of the system becomes less frequent due to larger interatomic distances.

As mentioned above, electrons with high energies from 3PI (yellow) are present at all excess energies when a BEC is exposed to the laser pulse. Depending on the excess energy, between 43% and 83% of the detected electrons originate from 3PI, contradicting the simplified picture in Fig. 1 which solely considers 2PI. This results in a high charge imbalance for all pulse wavelengths, even for vanishing (two-photon) excess energy. Thus, the fast 3PI electrons prevent the creation of a neutral plasma. For the dilute target, 3PI is reduced because significantly fewer atoms are located in the high-intensity volume of the femtosecond pulse and the plasma is more neutral. Again, this is caused by the overlap between the intensity distribution of the femtosecond laser beam and the atomic cloud (see "Methods"), where due to the small size and high density of the BEC many atoms are exposed to the high-intensity region close to the beam waist while in the larger thermal cloud, also low-intensity regions contribute to the electron signal.

Finally, we note that plasma electrons (green) recorded for positive and negative excess energies are expected due to the aforementioned microplasma formation triggered by direct photoionization of the atoms. For negative excess energies, the plasma originates from ionization of Rydberg atoms through Penning ionization and collisional ionization between free electrons and Rydberg atoms. We observe fewer plasma electrons for decreasing excess energy, since deeply bound states are harder to ionize, yet for a BEC, there is a significant fraction of electrons in the plasma even for very low excess energies.

For all measurements, we probe the final electron energies after the many-body dynamics taking place on pico- to nanosecond timescales. As the extraction fields of the detectors are constantly applied, the system evolves in a non-zero field configuration and the measured electrons are integrated over time. For the highly charged microplasma, this field configuration yields a lifetime of approximately 50 ns[20]. We expect similar lifetimes for the systems investigated here, since the total amount of charge carriers is comparable. The extraction fields (see Fig. S1 in Supplementary Note 1) are weak compared to the Coulomb interaction of the particles at the given densities, thus we do not expect a large perturbation of the dynamics. We also do not expect an influence of the extraction fields on the bound states as field ionization due to the extraction field strengths of $E_{ext} = 162$ V/m is only possible for Rydberg states with a principal quantum number exceeding $n = [Ry^2 \pi \epsilon_0 / (E_{ext} e^3)]^{1/4} = 119$ with $Ry = 13.6$ eV being the Rydberg energy, $\epsilon_0$ the dielectric field constant and $e$ the elementary charge.

## Simulations of the underlying dynamics

To understand the dynamics of the many-body systems at the microscopic level, we employ state-of-the-art molecular dynamics simulations using the *SARKAS - Python MD code for plasma physics*[53]. After the femtosecond laser pulse, the atoms in the ionization volume are modeled by ions and electrons that interact via the Coulomb potential in a field-free environment (see "Methods"). The initial ion distribution is approximately a cylindrical volume with a length of 5 µm and a diameter of 1.35 µm containing approximately 4200 ions (see "Methods"). The initial energy of the electrons corresponds to the energy resulting from 2PI or 3PI as described in the Methods section. In these simulations the cross section for 3PI is taken from our experimental data, since to the best of our knowledge no ionization cross sections are available at these intensities. The statistical uncertainty on the experimentally determined number of 3PI electrons is reflected by the shaded area around the free/3PI signal in Fig. 3a, representing the standard deviation of the signal.

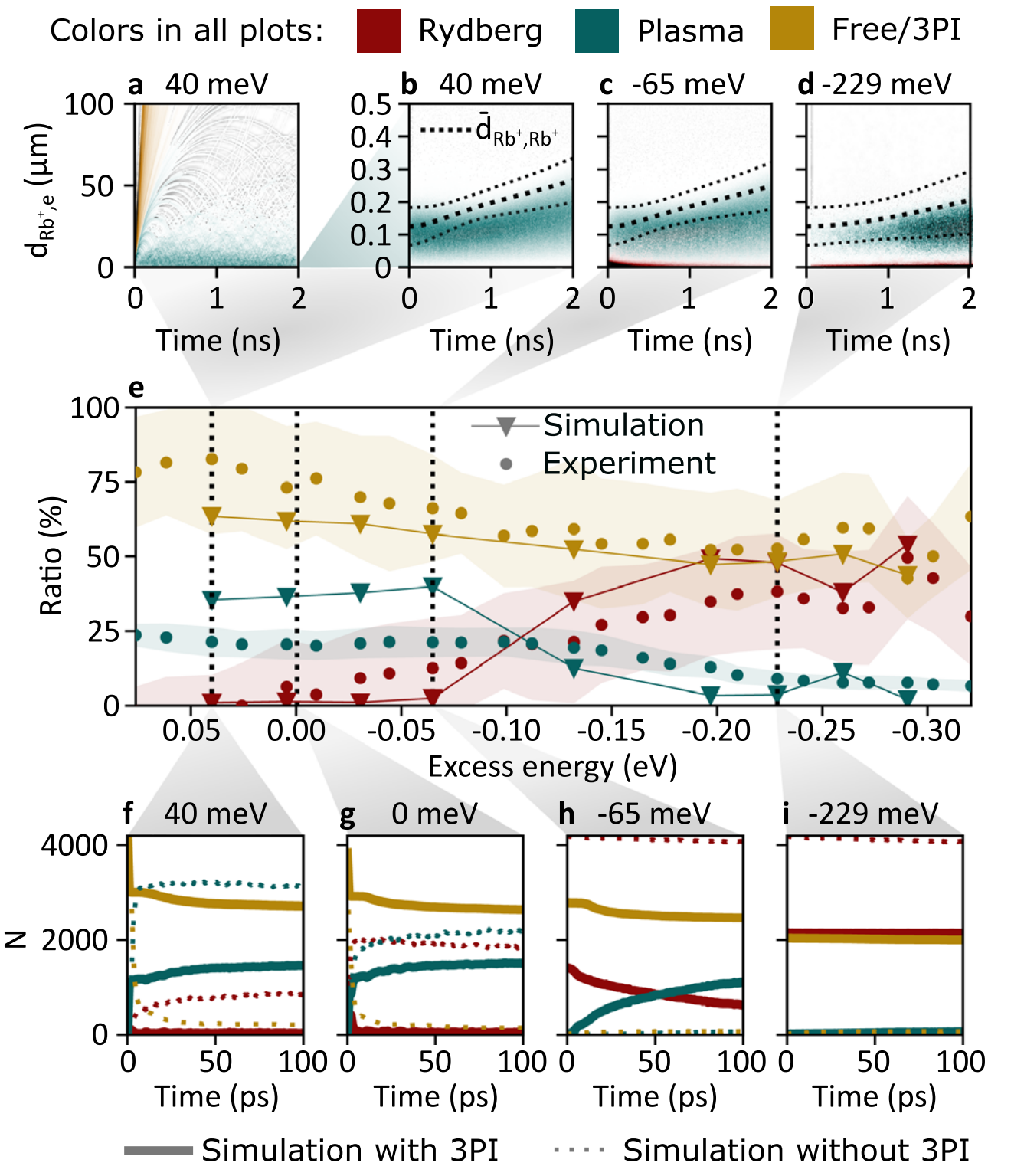

**Fig. 4 | Underlying dynamics of ultracold plasma and dense Rydberg gases.** All simulations are performed at a Bose-Einstein condensate (BEC) density of $\rho_{BEC} = 1.6 \times 10^{20}\ m^{-3}$. **a–d** Simulated electron trajectories referenced to the nearest ion $d_{Rb^+,e}$ over 2 ns. The electrons are either plasma electrons (green), Rydberg electrons (red) or electrons escaping the system (yellow). **a, b** At $E_{exc} = 40$ meV, free electrons quickly leave the system **a**. On smaller length scales (**b**) plasma electrons inside the ion cloud become evident as the electron-ion distance is shorter than the mean ion-ion distance $\bar{d}_{Rb^+,Rb^+}$. **c** At $E_{exc} = -65$ meV, Rydberg states are initially populated but quickly decay into plasma. **d** At $E_{exc} = -229$ meV, the population of Rydberg states remains stable throughout the simulated time frame. **e** Comparison between the ratio of Rydberg (red), plasma (green) and three-photon ionization (3PI)/free electrons (yellow) in the experiment (dots, shaded area indicates standard deviation) and simulation after 2 ns (triangles). **f–i** Simulations with (solid lines) and without 3PI (dotted lines) of the evolution of the number of electrons in Rydberg states (red), plasma electrons (green) and free electrons (yellow) within the first 100 ps. **f** At $E_{exc} = 40$ meV, the simulation without 3PI shows an almost neutral plasma with recombination of cold electrons into Rydberg states, while 3PI leads to the formation a highly charged microplasma. **g** At $E_{exc} = 0$ meV, 3PI also leads to the formation a highly charged microplasma. **h** At $E_{exc} = -65$ meV, the Rydberg population remains stable when neglecting 3PI, otherwise it quickly decays into plasma. **i** At $E_{exc} = -229$ meV, the Rydberg population remains stable within the simulated time frame for both cases.

A challenging aspect at high densities corresponding to a BEC ($\rho_{BEC} = 1.6 \times 10^{20}\ m^{-3}$) and low electron excess energies is that the binding energy becomes important. Thus, electrons and ions can not be randomly distributed, since this may lead to a Coulomb energy between an electron and the nearest ion neighbor being higher than the excess energy if pairwise binding energy is not taken into account. Instead, we create the electrons and ions pairwise in specific bound states reproducing the discrete energy spectrum of $^{87}$Rb or the excess energies corrected by the binding energy in the case of ionization.

Based on the simulation results, we can distinguish between the different classes of electrons by evaluating the potential and kinetic energies of the individual particles:

$$\begin{aligned} E_{k,e} \le E_{C,NN} &\Rightarrow \text{Rydberg electron} \\ E_{C,NN} < E_{k,e} \le E_{C,tot} &\Rightarrow \text{plasma electron} \\ E_{k,e} > E_{C,tot} &\Rightarrow \text{free electron} \end{aligned} \qquad (2)$$

Here, $E_{k,e}$ is the kinetic energy of the electron, $E_{C,NN}$ the potential energy the electron experiences from its nearest ion neighbor and $E_{C,tot}$ is the potential each electron experiences from the whole particle distribution (i.e. the attractive potential of all ions and the repulsive potential of all electrons). Thus, electrons in Rydberg states are electrons where the Coulomb energy of the nearest ion neighbor exceeds the kinetic energy, plasma electrons are electrons that are bound to the total Coulomb potential of all particles (but not to the nearest neighbor alone) and free electrons are electrons where the kinetic energy is larger than the total Coulomb energy.

To gain insights into the dynamics after the femtosecond laser pulse, Fig. 4a–d shows the distance between each electron and the nearest ion $d_{Rb^+,e}$ during the simulated time span of 2 ns at different excess energies. At $E_{exc} = 40$ meV (Fig. 4a, b) plasma electrons (green) are bound on large orbits around the ionic cloud, which is characteristic of the dynamics of a charged microplasma and the associated rapid cooling of the electrons. As expected, the electrons from 3PI immediately escape the system (yellow). By zooming in on shorter length scales (Fig. 4b), more plasma electrons can be seen inside the ionic cloud, where the ion-electron distance is smaller than the mean ion-ion distance $\bar{d}_{Rb^+,Rb^+}$ (bold dotted line, dotted line shows standard deviation). At $E_{exc} = -65$ meV (Fig. 4c), a fraction of electrons is initially bound in Rydberg states (red) which quickly decay into plasma electrons, while at $E_{exc} = -229$ meV (Fig. 4d), the Rydberg population remains constant throughout the simulation.

As shown in Fig. 4e, the ratios of Rydberg (red), plasma (green), and 3PI (yellow) between measurements (dots, shaded area shows standard deviation) and simulation results (triangles) show a very good agreement at all excess energies. In particular, the simulations show that a significant fraction of the initially created bound states survives the plasma dynamics at negative excess energies. The significantly smaller plasma fraction in the experiments close to the 2PI threshold can be explained by the extraction field of the detector setup, perturbing larger electron trajectories and saturation of the MCP detector at low electron energies, since most of the plasma electrons are spread over only a few channels of the detector (see low energy electrons in Fig. 2b).

The numerical simulations allow a direct comparison of the time-evolution of the electron composition taking into account only 2PI and Rydberg excitation (dotted lines) or including 3PI (solid lines) as depicted in Fig. 4f–i.

Above the threshold at $E_{exc} = 40$ meV (Fig. 4f), the simulations without 3PI show an almost neutral plasma in which only a small proportion of the electrons escape from the system (yellow), most of them are bound in the plasma dynamics (green). Even on these short time scales, recombination into Rydberg states (red) can be evidenced. In contrast, the simulation including 3PI shows a highly charged microplasma in which most of the electrons escape the system and only a few are bound in the plasma dynamics.

At the threshold (Fig. 4g), the simulation without 3PI shows a system closely resembling the idea discussed in Fig. 1c: Half of the electrons are ionized with excess energies corresponding to the bandwidth of the pulse and half of the electrons are excited to Rydberg states. The dynamics in the first picoseconds then balance between Rydberg recombination and ionization, with a slight preference for ionization. 3PI in turn leads to a highly charged microplasma in which only some of the electrons are bound in the plasma dynamics and no recombination takes place.

Below the threshold at $E_{exc} = -65$ meV (Fig. 4h), the simulation without 3PI shows an almost stable Rydberg ensemble within the simulated time frame, while the simulation including 3PI shows a fast transfer from Rydberg states into plasma due to charge imbalance. Thus, collisions with free charged particles due to 3PI are mainly driving the transition from excited Rydberg atoms into plasma. Only for low lying Rydberg states at $E_{exc} = -229$ meV (Fig. 4i), the Rydberg population remains stable for both cases in the time frame of 100 ps.

The very good agreement between simulations and experimental results without free parameters indicates that a classical model with Coulomb interaction between charged particles is sufficient to describe the many-body dynamics, including recombination into Rydberg states as well as the ionization mechanisms. For these simulations, it is crucial that the initial energies of the electrons are chosen correctly. The simulated time is short compared to the measured plasma lifetime for a similar system[20] but quantitatively reproduces all measurements, since the dynamics always take place within the first nanosecond due to the high density of the many-body systems. The reported dynamics is clearly different from that of an ultracold neutral plasma, where a back and forth between plasma and Rydberg gas due to ionization and recombination processes is expected[28,47]. These simulations also reveal the role played by 3PI processes during the femtosecond laser pulse in our experimental results, which leads to photoelectrons with high kinetic energy and free ions. The resulting high charge imbalance hinders recombination into Rydberg states and leads to ionization of even deeply bound Rydberg states.

## Discussion

We have investigated the many-body dynamics triggered by femtosecond laser pulses in an ultracold quantum gas leading to the formation of highly charged microplasma or dense Rydberg gases. Our experimental setup allows both, tuning the energy of the electrons as well as the system density over a wide range of initial parameters. The large bandwidth of the femtosecond laser pulse enables us to create dense ensembles of Rydberg atoms beyond the Rydberg blockade. Our detection scheme allows to directly probe the electron kinetic energy and distinguish between free, bound and plasma electrons.

Thus, we experimentally realize an excellent model system that allows comparison to exact molecular dynamics simulations, including Rydberg states since it comprises only a few thousand particles. These simulations model electrons as individual particles, and thus include collisional ionization and recombination processes in a comprehensive manner. Our model provides insight into the many-body dynamics on a microscopic level and thus transcend rate equations or hydrodynamic models focusing on macroscopic parameters. In the past, collisional processes leading to self-ionization in Rydberg gases have been studied both experimentally[37–40] and theoretically[41,42], yet a comprehensive and full-scale microscopic model of the many-body dynamics in ultracold plasma and dense Rydberg gases has not yet been implemented. Our experimental observations corroborated by molecular dynamics simulations both show different dynamics than expected from an ultracold neutral plasma where a back and forth between plasma and Rydberg gas by ionization and recombination processes is expected[28,47].

The obtained results now enable addressing the initial questions framing the investigation:

*How stable are dense Rydberg gases with overlapping electron orbits?*—Classically, as shown in the MD simulations, overlapping electron orbits are unstable. As soon as the orbits start to approach each other, Penning ionization will set in and the ionized fraction increases with the spatial overlap (see Fig. S2 in Supplementary Note 3). In our experiment overlap is only possible for $n > 45$, even in the dense case of an excited BEC with a Wigner-Seitz radius of approximately 100 nm. Quantum mechanically, a coherent superposition of overlapping electron wavefunctions may induce chemical bonds, however, such effects are experimentally not accessible in our system due to the charge imbalance always created by the 3PI electrons leading to ionizing electron-Rydberg collisions.

*Is it possible to create a plasma with negligible electron energy out of Rydberg excitations to reach the strong-coupling regime?*—Although our experimental setup allows tuning the excess energy to negative values, we always create additional charge imbalance due to 3PI electrons. In a system without additional free charge carriers, however, Penning ionization would always lead to finite excess energies of the released electrons. Even if negligible excess energies can be generated, e.g. by an ionization pulse with vanishing excess energy after Rydberg excitation, disorder-induced heating would prevent reaching the strong-coupling regime (see Fig. S4 in Supplementary Note 5) unless spatial correlations are already introduced in the atomic system (e.g., by using optical lattices, optical tweezers or a blockaded Rydberg gas).

*Is there a difference between an ionized plasma with vanishing excess energy and a plasma forming from a Rydberg gas with vanishing binding energy?*—Our experiment accesses this transition when crossing the two-photon ionization threshold so that the electron excess energy changes sign. However, we do not see a particularly different dynamics slightly above or below $E_{exc} = 0$. Figure 4g shows the case for vanishing excess energy and indicates a rapid decay of the created bound state due to the 3PI electrons in the experiment (solid lines). However, the MD simulations show a considerably slower decay of bound states into plasma in the absence of 3PI electrons (dashed lines).

*Is there an equilibrium between three-body-recombination and ionizing collisions?*—Our experimental observations as well as MD simulations always show a preference towards either three-body-recombination (dashed red line in Fig. 4f) or ionizing collisions (all other red lines in Fig. 4g–i). An investigation on the existence of specific parameters in terms of density, excess energy distribution and excited state fraction that balance three-body-recombination and ionizing collisions is beyond the scope of this work.

In summary, the very good agreement between simulations and experimental results suggests that a classical model is sufficient to describe the dynamics of our many-body system. Due to the high intensity of the femtosecond laser pulse, 3PI processes lead to an initial charge imbalance, which hinders the three-body recombination of low-energy electrons into Rydberg states and limits the stability of the investigated Rydberg gases. These higher-order ionization processes could be suppressed by using an ultraviolet (UV) laser pulse for single-photon ionization, which would significantly reduce the required intensity.

Further insight is expected by employing complete ionization experiments using, e.g., ion microscopy[54,55] and velocity-map-imaging[56–58] techniques with coincident detection of the ions and electrons[59]. This makes it possible to distinguish whether an excited Rydberg state is generated directly during the excitation pulse or arises in the subsequent plasma dynamics. A time-resolved measurement provides further insights into the dynamics of the many-body system and the stability of dense Rydberg gases, as it enables direct access to the time-evolution of Penning ionization and three-body recombination products.

Photoionization of ultracold quantum gases constitutes a promising platform to realize strongly coupled plasma as they allow for strong initial ion coupling. In addition, starting from a 3D Mott insulator in optical lattices, disorder-induced heating should be suppressed, allowing to investigate whether a strong spatial correlation in a plasma also can conserve initially strong coupling[60]. Ultrashort laser pulses allow creating Rydberg excitations beyond the Rydberg blockade in dense atomic gases. Pico- and femtosecond laser pulses provide a promising tool to address Rydberg states via ultrafast Rabi oscillations[61] and to induce ultrafast energy exchange between Rydberg states[62]. This is an important step forward towards realizing ultrafast gates for quantum simulation and quantum computing operating at the speed-limit of the dipole-dipole interaction. However, a concrete realization in dense atomic ensembles depends on the exact excitation and ionization dynamics during the ultrashort laser pulse, demanding further studies for different atomic arrangements and pulse shapes.

Combining ultracold quantum gases with ultrashort laser pulses should allow investigating if a transient metal-like phase emerges, where the electron wavefunction is shared by several atoms yielding an electronic band structure[18,19]. 1D optical lattices extend the range of addressable targets to electronic systems with reduced dimensions similar to 2D-systems in solid-state materials[63,64] or arbitrary configuration by using optical tweezer arrays[65–67].

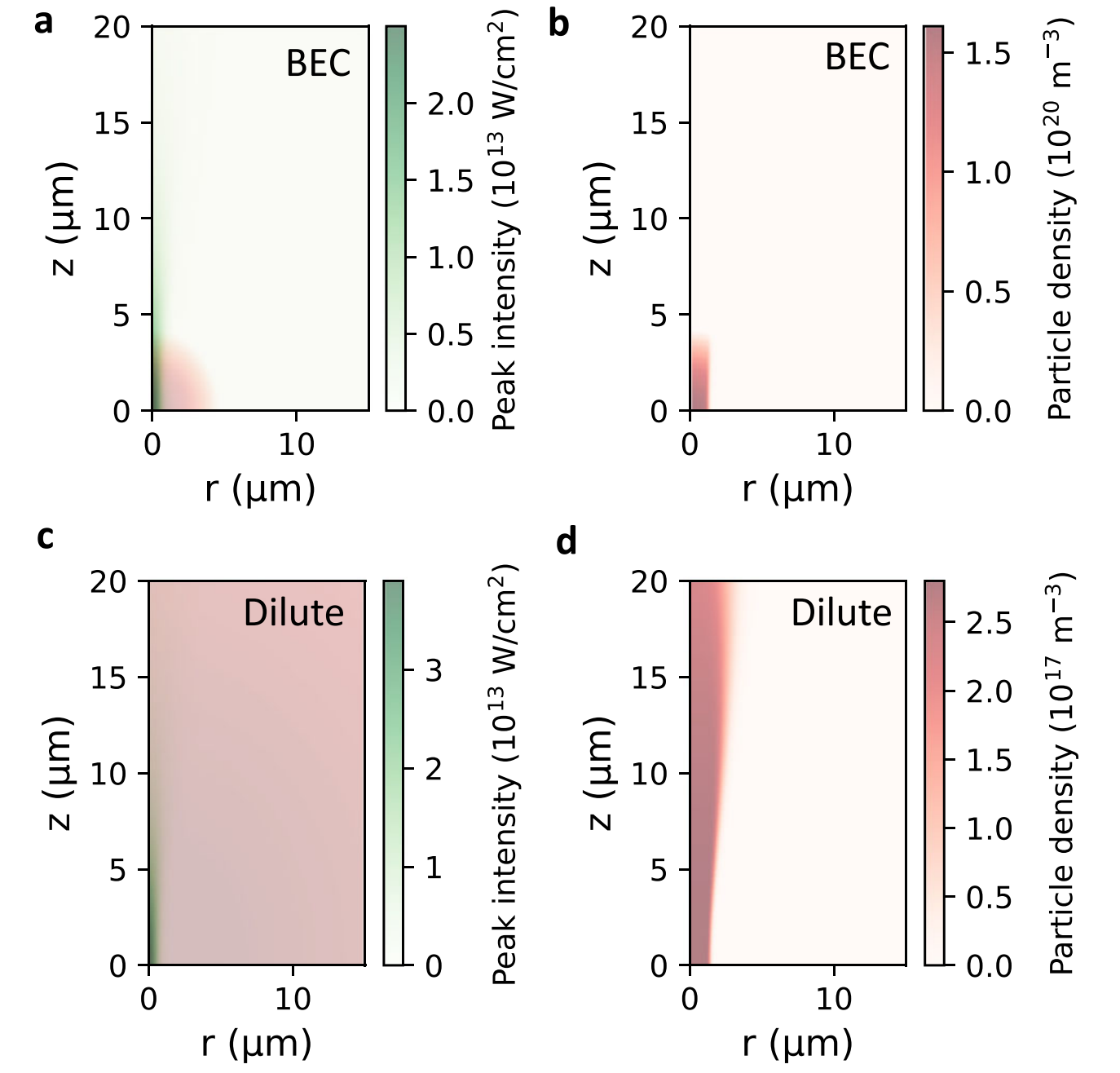

**Fig. 5 | Spatial peak intensity and atom/ion density distribution.** The overlap between the spatial intensity distribution of the Gaussian femtosecond laser beam with a waist of $w_0 = 1$ μm with the atomic density of a Bose-Einstein condensate (BEC) (**a**) given by a Thomas-Fermi distribution and a thermal cloud (**c**) given by a Gaussian distribution determines the initial distribution of ions and atoms in excited states. The small size of the BEC leads to a compact initial ion distribution (**b**) compared to the large thermal cloud where the atomic density extends further into the low intensity regions along the propagation direction $z$ of the femtosecond laser beam (**d**).

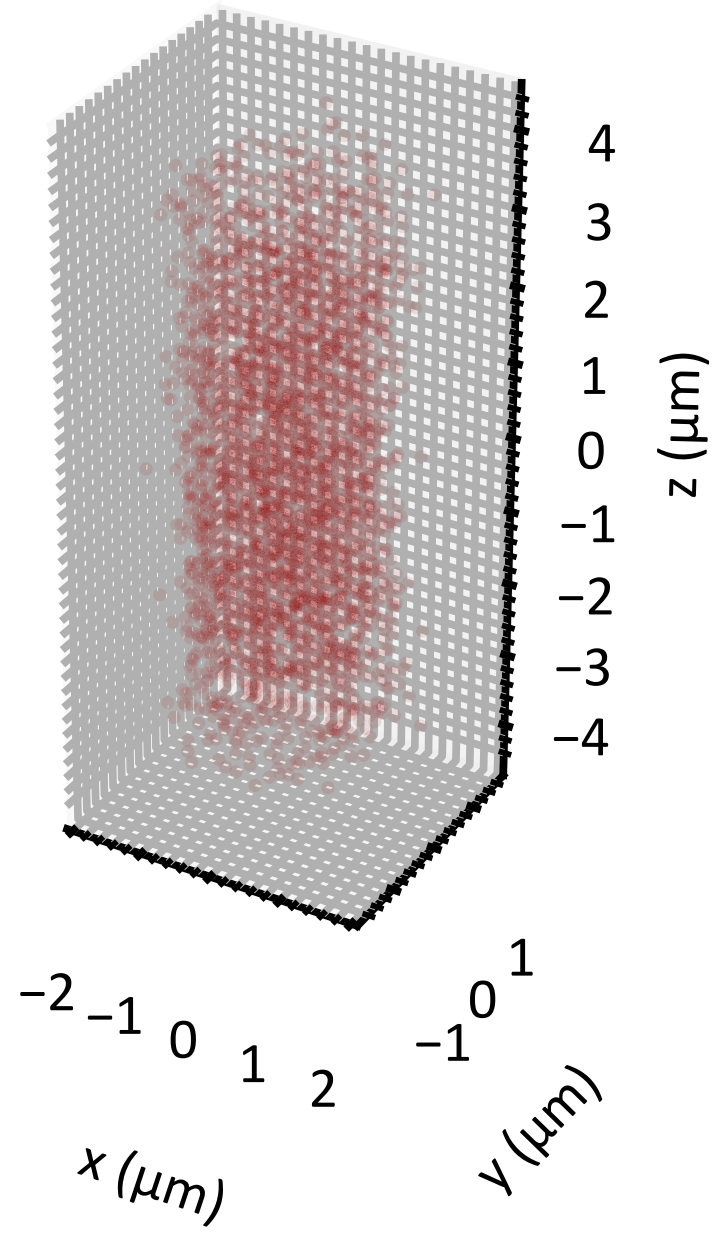

**Fig. 6 | Ion distribution used for molecular dynamics simulations.** Calculated distribution of 4193 ions created by exposing a Bose-Einstein condensate (BEC) to a single femtosecond laser pulse of 590 nm.

## Methods

### Ultracold atoms

We use alkali metal dispensers to generate a dilute $^{87}$Rb vapor in a glass cell. The atoms are first collected in a 2D magneto-optical trap (MOT) to improve the loading of our 3D MOT separated by a differential pumping stage. A short gray molasses phase after the 3D MOT allows us to reach temperatures on the order of a few μK.

After laser cooling, we load the atomic ensemble into a hybrid trap consisting of a magnetic quadrupole field and a dipole trap at 1064 nm and employ forced radio-frequency evaporation to further increase the phase-space-density of the cloud. To move the atomic cloud into a field-free environment, we use the beam of the optical dipole trap as an optical tweezer and transport the atoms into a second vacuum chamber, containing the charged particle detection, by moving the focus of the beam. Here, we shine in a second dipole trap beam and evaporate the atoms to quantum degeneracy in a crossed dipole trap with final trap frequencies $\omega_{x,y} = 2\pi \times 113(3)$ Hz and $\omega_z = 2\pi \times 128(1)$ Hz.

All measurements discussed are performed with either a BEC with a peak density of $\rho_{\text{BEC}} = 1.6 \times 10^{20}$ m$^{-3}$ or a dilute thermal cloud with a peak density of $\rho_{\text{th}} = 2.8 \times 10^{17}$ m$^{-3}$.

### Femtosecond laser pulses

We use a commercially available ytterbium-doped potassium-gadolinium tungstate (Yb:KGW) solid-state laser system to generate the femtosecond laser pulses via Kerr-lens mode-locking and chirped-pulse amplification (CPA) in a regenerative amplifier (RA). The pulses with a pulse duration of 300 fs FWHM, a fundamental wavelength of 1022 nm and a bandwidth of 4.36 nm FWHM are fed into an OPA.

Here, second harmonic generation and difference frequency generation is used to generate pulses in the range between 510 and 710 nm with a duration of 166(3) fs. The duration is determined by measuring the bandwidth of 3.1(1) nm FWHM at a central wavelength of 593.7 nm and assuming a Fourier-limited Gaussian temporal pulse profile. With a Pockels cell based pulse picker we are able to select single pulses from the pulse train in the RA to create single pulses in the OPA for the presented measurements.

A microscope objective with a numerical aperture of 0.5 focuses the pulses onto the atoms with a waist of $w_0 = 1$ μm. All measurements with a BEC are performed at pulse peak intensities of $I_{0,\text{BEC}} = 2.5(2) \times 10^{13}$ W/cm$^2$ while all measurements with a dilute thermal cloud are performed at pulse peak intensities of $I_{0,\text{th}} = 3.9(2) \times 10^{13}$ W/cm$^2$.

### Density and intensity overlap

The initial spatial distribution of ionized or excited particles depends on the overlap of the atomic target (BEC or dilute thermal cloud) with the spatial intensity distribution of the femtosecond laser beam. For all experiments the waist of the femtosecond laser beam of $w_0 = 1$ μm is smaller than the atomic cloud, so that the excitation or ionization is localized in the center of the cloud.

Figure 5 exemplifies the spatial overlap between the intensity distribution of the femtosecond laser beam with the atomic density distribution for a BEC (Fig. 5a) and a dilute cloud (Fig. 5c).

The local peak intensities $I_0(r, z)$ translate to a spatial map of ionization probabilities $P(I_0(r, z))$ that give the initial ion density $\rho_{\text{ion}}(r, z) = \rho_{\text{atom}}(r, z) \times P(I_0(r, z))$ by multiplication with the local atom density $\rho_{\text{atom}}(r, z)$.

In both cases, the radial intensity profile of the femtosecond laser beam is much smaller than the atomic density distribution. However, in propagation direction $z$ of the laser beam, the Thomas-Fermi profile of the BEC

$$\rho_{\text{TF}}(r, z) = \rho_{\text{BEC}} \times \max\left[0, 1 - \left(\frac{r}{R_{\text{TF}}^r}\right)^2 - \left(\frac{z}{R_{\text{TF}}^z}\right)^2\right] \quad (3)$$

with peak density $\rho_{\text{BEC}} = 1.6 \times 10^{20}$ m$^{-3}$ and Thomas-Fermi radii $R_{\text{TF}}^r = 4.7$ μm and $R_{\text{TF}}^z = 4.2$ μm clips the areas of high peak intensities so that the initial ion distribution is determined by the waist $w_0$ of the focused laser beam along $r$ and the Thomas-Fermi radius of the BEC along $z$ (Fig. 5b). This leads to the cylindrical initial ion distribution in Fig. 6.

For a dilute, thermal cloud with an Gaussian density profile

$$\rho_{\text{Gauß}}(r, z) = \rho_{\text{th}} \times \exp\left(-\frac{r^2}{2\sigma_r^2} - \frac{z^2}{2\sigma_z^2}\right) \quad (4)$$

including peak density $\rho_{\text{th}} = 2.8 \times 10^{17}\,\text{m}^{-3}$ and an rms width of $\sigma_r = \sigma_z = 35.2\,\mu\text{m}$ the spatial extent along $z$ is much larger and also reaches into the low intensity regions of the laser beam along propagation direction so that the initial ion profile is more elongated along $z$ (Fig. 5d). However, as the ionization probabilities in the focal region are close to $P = 1$, the peak ion density also is much lower (compare colorbars in Fig. 5b, d).

## Rydberg detection

Atoms in bound states, including Rydberg atoms are detected by recording the photoelectrons after ionization via a 10 µs pulse at $\lambda_{\text{Ry,PI}} = 1064$ nm at an intensity of $I \approx 7 \times 10^7\,\text{W/cm}^2$.

We assume that the applied pulse ionizes most of the Rydberg atoms created by the femtosecond laser pulse, since the Rydberg electron signal saturates with increasing power of this Rydberg ionization pulse. Theory suggests, that $s$-states are more unlikely to be ionized. However, the calculated cross sections given in ref. 68 cover only states with $n \geq 20$, for lower lying states a higher cross section seems to be plausible. Figure 7 shows the calculated ionization probabilities for the 10 µs pulse used in the experiment. The extracted electron numbers from the measurement in Fig. 3a, thus are only a lower bound for the actual number of Rydberg atoms.

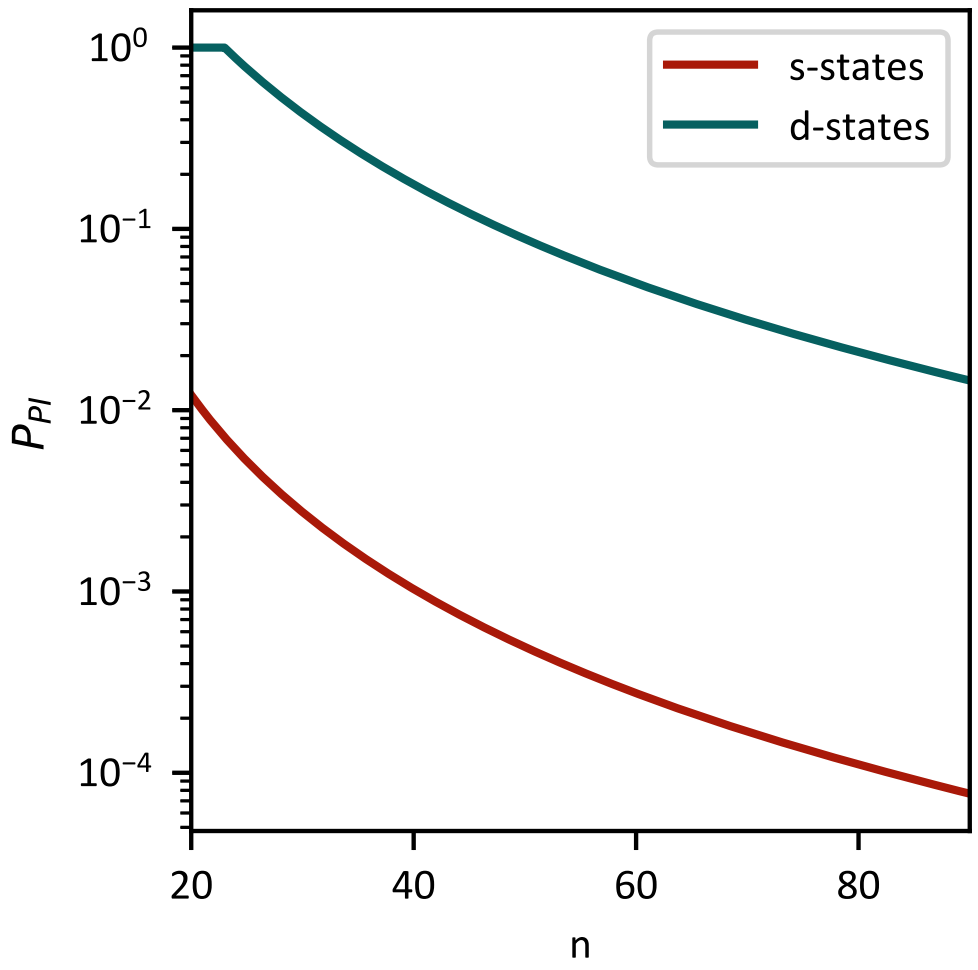

**Fig. 7 | Ionization probabilities of Rydberg *s* and *d*-states.** Calculated using the cross sections from ref. 68 and assuming a 10 µs pulse at $\lambda_{\text{Ry,PI}} = 1064$ nm and an intensity of $I \approx 7 \times 10^7\,\text{W/cm}^2$.

## Molecular-dynamics simulations

In the molecular dynamics simulations described in the Results section of the main manuscript we use open boundary conditions. We implement the Coulomb force with a short-range cutoff:

$$F_{\text{C}} = \begin{cases} \frac{1}{4\pi\epsilon_0}\frac{q_1 q_2}{r^2} & , r > R_0 \\ \frac{1}{4\pi\epsilon_0}\frac{q_1 q_2}{R_0^2} & , r \leq R_0 \end{cases} \quad (5)$$

to avoid the divergence at small distances. Here, $\epsilon_0$ is the vacuum permittivity, $q_1$ and $q_2$ are the charges of the interacting particles, $r$ denotes the distance between these particles and $R_0$ is the cutoff distance, here set to 2.5 nm to cover binding energies down to $E_{\text{b}} \approx -0.58$ eV. This binding energy corresponds to the 8$s$ state and is well below the smallest experimentally accessed excess energy of $E_{\text{exc}} = -0.32$ eV (corresponding to the 10$s$ state). The simulation code uses a velocity-Verlet integrator and the time step was set to 1 fs to ensure energy conservation.

The spatial ion distribution is calculated by overlaying a Thomas-Fermi profile for the given trap parameters and density with a Gaussian beam intensity profile. The Thomas-Fermi profile of the cloud is derived from a BEC with $N = 2.5 \times 10^4$ atoms in a crossed dipole trap with trap frequencies of $\omega_{x,y} = 2\pi \times 113$ Hz and $\omega_z = 2\pi \times 128$ Hz yielding a peak density of $\rho_{\text{BEC}} = 1.6 \times 10^{20}\,\text{m}^{-3}$.

The femtosecond laser pulse was included with a pulse duration of 166 fs at 590 nm with a waist of $w_0 = 1\,\mu$m. At a pulse energy of 60 nJ this yields a pulse with a peak intensity of $I_0 = 2.2 \times 10^{13}\,\text{W/cm}^2$. We use the 2PI cross sections measured on ultracold $^{87}$Rb in a MOT setup reported in ref. 69.

The resulting initial ion distribution is a cylindrical volume with a total length of 5 µm and a diameter of 1.35 µm containing approximately 4200 ions (see Fig. 6), which is used for all simulations presented here. While the geometry of the initial spatial configuration does play a role for the dynamics and leads to, e.g., anisotropic plasma frequencies of the electrons, the exact configuration and placement of the individual ions for a given density is not relevant for the number of involved particles in our simulation due to the dominance of disorder induced heating.

At high densities and low electron excess energies, electrons and ions can not be randomly distributed since the Coulomb energy between an electron and the nearest ion neighbor can be higher than the excess energy. Instead, we generate the electrons and ions pairwise and take the initial binding energy of the electron to its paired ion into account. This also allows us to create specific bound states between electrons and ions and energetically reproduce the expected energy spectrum when we excite with a pulse at wavelengths below the 2PI-threshold.

As initial ion-electron distance we choose a distance of 3.8 nm to be able to model deeply bound states. The Coulomb energy at this electron-ion distance has to be carefully considered: to generate an electron with an excess energy of $E_{\text{exc}}$ we have to give the electron a kinetic energy of $E_{\text{kin}} = E_{\text{exc}} - E_{\text{C,i}}$, where $E_{\text{C,i}} < 0$ is the Coulomb energy of the initial distance.

**Fig. 8 | Exemplary probability distribution of electron energies.** Probability distribution of electron energies at a wavelength of 593.7 nm and approximately 50% three-photon ionization (3PI) events. While the two-photon process yields bound states and free electrons close to the threshold (inset), the three-photon process yields free electrons with a kinetic energy of $E_{\text{exc}} = 2.09$ eV.

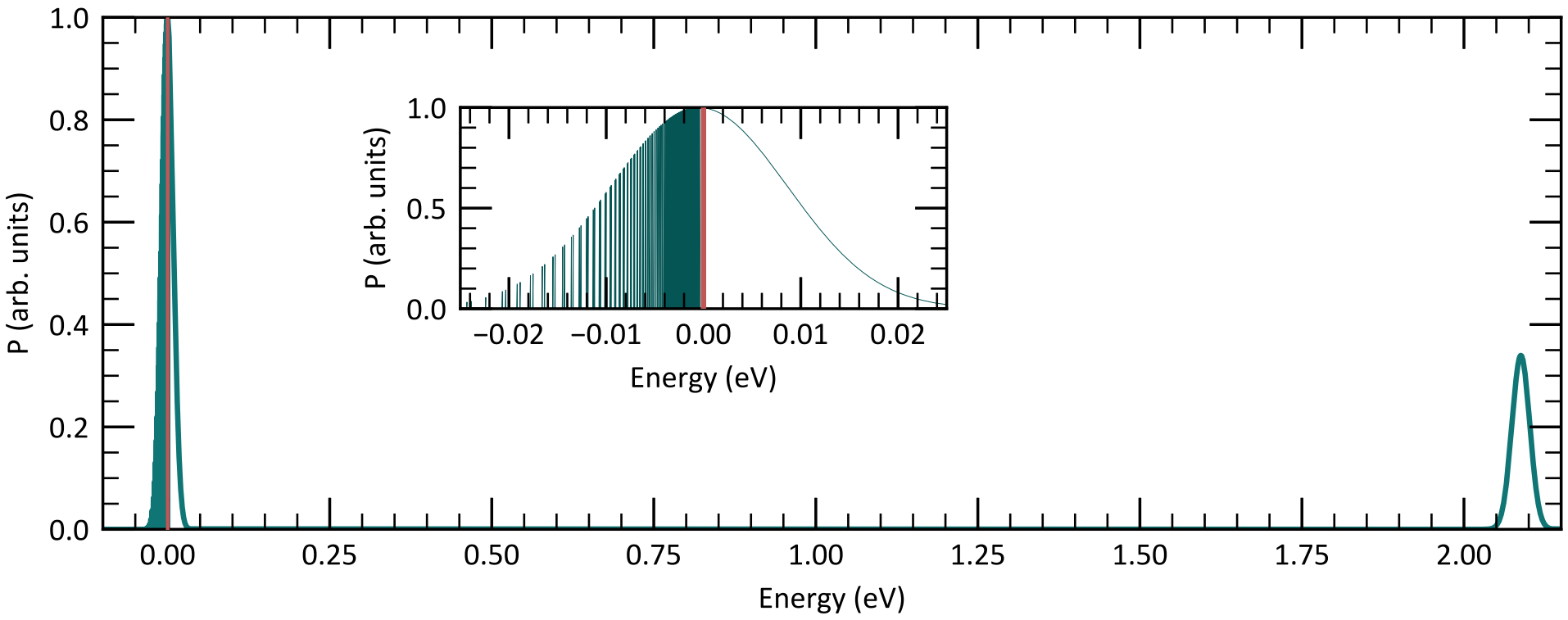

Positive excess energies yield ionization processes, while negative excess energies yield electrons bound in Rydberg states.

Since the excitation probabilities below the threshold are, to the best of our knowledge, not easily calculated in the strong-field regime, we simply scale the excitation probability with the spectral intensity of the pulse taking only the *s* and *d* states of the unperturbed energy spectrum of $^{87}$Rb[70] into account, which reflects the selection rules of the two-photon process. The ratio of 3PI events was set to the experimentally determined ratios (see Fig. 4e).

Figure 8 shows an electron spectrum generated this way at a wavelength of 593.7 nm right at the 2PI-threshold, including 50% 3PI events with an excess energy of $E_{exc}$ = 2.09 eV. The large bandwidth of the laser pulse covers various states below the 2PI-threshold (Fig. 8 inset) and also reaches into the continuum above the threshold.

## Data availability

The data that support the findings of this study are available from the corresponding author upon request.

## Code availability

Any custom computer code used in this study is available from the corresponding author upon request.

## Acknowledgements

This work is funded by the Cluster of Excellence "Advanced Imaging of Matter" of the Deutsche Forschungsgemeinschaft (DFG) - EXC 2056 - project ID 390715994. We acknowledge financial support from the Open Access Publication Fund of Universität Hamburg.

## Author contributions

The experiment was coordinated by M.D., K.S., P.W.S., and J.S. Setup, measurements, and data analysis were performed by M.G. with support from J.F., J.H., P.W.S., J.S. and discussions involving all co-authors. The molecular dynamics simulations were performed by M.G. with support from P.W.S. The manuscript was written by M.G. with the support of J.H. and P.W.S. and discussions involving all co-authors.

## Funding

Open Access funding enabled and organized by Projekt DEAL.

## Competing interests

The authors declare no competing interests.

## Additional information

**Supplementary information** The online version contains supplementary material available at https://doi.org/10.1038/s42005-026-02674-9.

**Correspondence** and requests for materials should be addressed to Philipp Wessels-Staarmann.

**Peer review information** *Communications Physics* thanks the anonymous reviewers for their contribution to the peer review of this work. A peer review file is available.